\documentstyle[sprocl]{article}
\input{epsfig.sty}
\def\Journal#1#2#3#4{{#1} {\bf #2}, #3 (#4)}

\def\npb{{\em Nucl. Phys.} B}
\def\plb{{\em Phys. Lett.}  B}
\def\prl{\em Phys. Rev. Lett.}
\def\prd{{\em Phys. Rev.} D}
\def\zpc{{\em Z. Phys.} C}

\def\to{\rightarrow}

\def\beq{\begin{equation}}
\def\eeq{\end{equation}}
\def\bea{\begin{eqnarray}}
\def\eea{\end{eqnarray}}
\def\beqa{\begin{eqnarray}}
\def\eeqa{\end{eqnarray}}
\def\etal{{\it{et al}}}

\def\ifmath#1{\relax\ifmmode#1\else$#1$\fi}
\def\etal{{\it et al.,}}

\def\to{\rightarrow}

\def\BR{\mbox{{\rm BR}}}

\def\gsim{{~\raise.15em\hbox{$>$}\kern-.85em
          \lower.35em\hbox{$\sim$}~}}
\def\lsim{{~\raise.15em\hbox{$<$}\kern-.85em
          \lower.35em\hbox{$\sim$}~}}
\def\kev  {\ifmath{\mbox{\,ke\kern -0.08em V}}} 
\def\mev  {\ifmath{\mbox{\,Me\kern -0.08em V}}} 
\def\gev  {\ifmath{\mbox{\,Ge\kern -0.08em V}}} 
\def\gevc {\ifmath{\mbox{\,Ge\kern -0.08em V$\!/c$}}} 
\def\mevc {\ifmath{\mbox{\,Me\kern -0.08em V$\!/c$}}} 
\def\gevcc{\ifmath{\mbox{\,Ge\kern -0.08em V$\!/c^2$}}} 
\def\mevcc{\ifmath{\mbox{\,Me\kern -0.08em V$\!/c^2$}}} 

\begin{document}

\begin{titlepage}

\begin{flushright}
UCTP 107.98
\end{flushright}

\vspace{1.5cm}

\begin{center}
\Large\bf 
The Phenomenology of Enhanced \boldmath$b\to sg$\unboldmath
\end{center}

\vspace{1.2cm}

\begin{center}
Alex Kagan\\
{\sl Department of Physics, University of Cincinnati\\
Cincinnati, OH 45221}
\end{center}

\vspace{1.3cm}

\begin{center}
{\bf Abstract}\\[0.3cm]
\parbox{11cm}{
Potential hints for New Physics enhanced $b \to sg$ dipole operators are reviewed.  
Implications for inclusive kaon spectra, and corrresponding search 
strategies are discussed.  Remarkably, all $B$ meson rare decay constraints can be evaded.
Large CP asymmetries are expected if new contributions to the dipole operators
contain non-trivial weak phases.  A critical comparison of $\eta'$ production in the Standard Model
and in models with enhanced $b \to sg$ is presented.  
A Standard Model explanation of the large $B \to \eta' X_s$
rate measured by CLEO poses a challenge, evidently requiring novel non-perturbative 
order of magnitude enhancement of gluon anomaly mediated 
contributions.
However, in the case of enhanced $b \to sg$ 
the existence of a cocktail solution is very likely.}
\end{center}

\vspace{1cm}

\begin{center}
{\sl To appear in the Proceedings of the\\
Seventh International Symposium on Heavy Flavor Physics\\
Santa Barbara, California, July 7-11 1997}
\end{center}

\end{titlepage}

\setcounter{page}{1}


\section{Introduction}

Enhanced $b \to sg$ refers to the possibility that
new physics enhances the $\Delta B =1 $ chromomagnetic dipole
operators so that $\BR(b \to sg) \sim 10\%$.
This is in contrast to the Standard Model predictions of 
$\BR(b \to sg) \sim .2\% $ for ``on-shell" glue~\cite{ciuchiniold},
and $\BR(B \to X_{no~charm} ) \sim 1 - 2 \% $~\cite{nierste}.
We begin with an updated account of potential hints 
for enhanced $b \to sg$ 
from inclusive $B$ decays~\cite{hougradzskowski}$^-$\cite{kaoncharm}, then discuss implications 
of such a scenario for 
the inclusive $K$ momentum spectra and for exclusive rare $B$ decays. 
The possibility of large $CP$
violating asymmetries in rare decays is emphasized.  
The last section is devoted to a comparison of 
$B \to \eta' K $ and $B \to \eta' X_s $ in the Standard Model 
and in models with enhanced $b \to sg$, in light of the large rates 
for these processes reported by the CLEO Collaboration.  

The theoretical motivation for enhanced $b \to sg$ follows from the
observation that it is often associated with generation of particular 
combinations of quark masses or CKM mixing angles 
via new dynamics at TeV scales~\cite{kagan}. 
The chirality flip inherent in new contributions to the dipole operators and 
quark mass matrices has a common origin, leading to direct correlations between the two.  
There are several known examples in which this connection 
can be realized without violating FCNC bounds: 
radiatively induced quark masses at one-loop
via exchange of gluinos and squarks, or via exchange of 
new charge -1/3 vectorlike quarks and neutral scalars, 
and dynamically generated quark masses in technicolor models with techniscalars.
Constraints from $\BR(b \to s \gamma)$ rule out~\cite{kagan,hougeng} enhanced $b \to sg$
via one-loop diagrams containing a top quark and charged Higgs.
The possibility of a large rate for $b \to sg$ in supersymmetric models was first 
noted in Ref.~\cite{masiero}.
A detailed discussion of $b \to s \gamma $ 
in supersymmetric models of enhanced $b \to sg$ can be found in 
Ref.~\cite{ciuchininew,kaganneubertCP}.  
It is also worth mentioning that models of quark substructure with order TeV
compositeness scales would be potential candidates since in this case gluon emmision by an exchanged 
preon participating in quark mass generation might also lead to significant dipole operator
contributions.  

\section{Hints for enhanced $b \to sg$ }

\subsection{Charm counting}

Some phenomenological consequences of enhanced $b \to sg$ for $B$ decays
are a decrease in the average charm multiplicity and semileptonic 
branching ratio~\cite{hougradzskowski}, and an increase
in the kaon yields~\cite{kaoncharm}.
Hints for all three are summarized below.  
More details are given in~\cite{kaoncharm,hawaii}.
 
Updated inclusive $B$ to charmed hadron flavor blind branching ratios 
used to obtain the $B$ decay charm multiplicity at the $\Upsilon(4S)$ are given
in Table~\ref{tab:one}.
For the $D/D_s $ yields we have averaged 
the ARGUS, CLEO 1.5 and CLEO
meaurements~\cite{argusD,cleo15D,cleonewD}.
The $D^0$ and $D^+$ yields have been rescaled to the world averages 
determined in Ref.~\cite{richman},
$\BR(D^0 \to  K^- \pi^+ ) = 3.87 \pm .09 \%$ and 
$\BR(D^+ \to K^- \pi^+ \pi^- ) = 8.8 \pm .6 \%$, respectively.\footnote{
In the former we 
have added the recent CLEO 
measurement $\BR(D^0 \to  K^- \pi^+ ) = 3.81 \pm .22 \%$,
obtained from partial 
reconstruction of $\overline{B} \to D^{*+} X \ell^- \bar {\nu}$ \cite{CLEODKPI}.}
The $D_s$ yields correspond to the PDG average~\cite{pdg96} $\BR(D_s \to \phi \pi) =
3.6 \pm  .9 \%  $. 
The charmed baryon
and charmonium yields are those recently used by
the CLEO collaboration~\cite{cleonewD,persis}.
The resulting world-average $B$ decay charm multiplicity at the $\Upsilon(4S)$ is 
\begin{equation}
 n^{exp}_c = 108 \pm 4.8 \% .\label{eq:nclow} 
\end{equation} 
Using only the recent CLEO $D/D_s $ yields~\cite{cleonewD} 
rather than the world averages gives $n_c = 112 \pm 5.3 \% $.
We will see that next-to-leading order Standard Model QCD predictions
are somewhat larger. 

\begin{table}[t]
\caption{
Inclusive $B \to~charmed~hadron$, and $B \to K$ branching ratios [$\%$]. 
$(c \bar c)$ is any $c \bar c$  meson.\label{tab:one}}
\vspace{0.4cm} 
\begin{center}
\begin{tabular} {| c | c | } \hline
Process  & Branching Ratio  \\
\hline
$\overline{B} \to D^0 / \overline{D^0} X $ & $62.1 \pm 2.5  $  \\
$\overline{B} \to D^+ /  D^- X $ &  $24.7 \pm 2.0 $ \\
$\overline{B} \to D_s^+  / D_s^- X $ &  $9.9 \pm 2.6 $     \\
$\overline{B} \to (c \bar c) X_s $ 
   & $ 2.7 \pm .35  $ \\
$\overline{B} \to \Lambda_c^+  / \Lambda_c ^- X $ & $ 3.9 \pm 2.0 $ \\
$\overline{B} \to \Xi_c^+  X $ & $ 0.8 \pm 0.5 $ \\
$\overline{B} \to \Xi_c^0  X $ & $ 1.2 \pm 0.9 $ \\
\hline
\end{tabular}
\end{center}
\end{table}

The flavor specific charmed hadron branching ratios 
in Table~\ref{tab:two} are obtained by combining the 
relative flavor specific yields with the corresponding
flavor blind yields in Table~\ref{tab:one}. The 
charmonium yield is included to obtain the $\Upsilon(4S)$
world-average
\beq \BR^{exp} (\overline{B} \to X_{c \bar c  s}) = 19.4 \pm 3.5 \%.\eeq
Using only the recent CLEO $D/D_s$ flavor blind yields rather than the world averages
gives $\BR (\overline{B} \to X_{c \bar c s}) = 21.2 \pm 3.6 \% $.
We'll see that this is consistent with next-to-leading order predictions.
$\BR(\overline{B} \to X_{c \bar u d}) $
can also be determined purely experimentally by combining flavor blind
and flavor specific charmed hadron yields~\cite{kaoncharm,hawaii}, giving
\beq \BR^{exp}(B \to X_{c \bar u d} ) =  45.7 \pm 6.6\%, \label{eq:Xcud} \eeq
also consistent with next-to-leading order predictions.
The result using only the CLEO $D/D_s $ yields is essentially the same.
Similarly, one finds
\beq  \BR(\overline{B} \to X_{c \bar u d} \to DX/D_s X) = 41.0 \pm 6.2\%,
\label{eq:Dcud} \eeq
which is used to normalize estimates of kaon production from $s \bar s$ popping.

\begin{table}[t]
\caption{
Inclusive flavor tagged $B$ decay branching ratios [\%]. 
$D$ denotes $D^0$ or $D^+$, and similarly for $\overline{D}$.\label{tab:two} }
\vspace{0.4cm} 
\begin{center}
\begin{tabular} {| c | c | c | c |} \hline
$T$  & ${{\cal B}(\overline{B} \to \overline{T}X) \over  {\cal B}(\overline{B}
\to TX) } $ &    
${\cal B}(\overline{B} \to \overline{T}X)$  &   ${\cal B}(\overline{B} \to TX) $
\\
\hline
$D$ &  $.100 \pm  .031$~\cite{CLEOBDDK}   & $7.9 \pm 2.2$      & $78.8 \pm 3.7 $ \\
$D_s^- $ & $.21 \pm .10 $~\cite{CLEODstag}  & $1.7 \pm .8$ & $8.2 \pm 2.6$ \\
$\Lambda^+_c $ &  $.19 \pm .14 $~\cite{CLEOLambdatagNEW}      &    $.6 \pm .5 $ & 
$3.3 \pm 1.7 $   \\
\hline
\end{tabular}
\end{center}
\end{table}

The charm multiplicity and $\BR(\overline{B} \to X_{c \bar c s})$ can be 
used to bound $\BR (\overline{B} \to X_{sg})$ via
the relation
\beq n_c = 1 + \BR(B \to X_{c \bar c s} ) - \BR(B \to X_{no~charm}).
\label{eq:ncBccs}\eeq
The above determinations of $n_c$ and $\BR(B \to X_{c \bar c s} )$
give                
\beq \BR (B \to X_{\rm{no~charm}}) = 11.4 \pm 5.9\%.    \label{eq:bxg1} 
\eeq
Using only the recent CLEO $D/D_s$ yields gives $9.3 \pm 6.4\%$.
For comparison, a recent NLO analysis~\cite{nierste} gives 
$\BR (B \to X_{\rm{no~charm}}) = 1.5 \pm .8 \% $ in the Standard Model.
Bounds on $\BR(B \to  X_{sg})$ follow by subtracting $\approx 1 \%$ to
account for $b \to u$ transitions.  
This is a potential hint for enhanced $b \to sg$, although it is also consistent 
with no $b \to sg$ at the $2 \sigma $ level.  
In the error bars approximately $\pm 3.4 \% $ is due to uncertainties 
in the $D$ decay branching fractions, i.e., $D^0 \to K^- \pi^+$, 
$D^+ \to K^+ \pi^- \pi^+ $, and $D_s \to \phi \pi $.  The remainder will
be considerably reduced at the $B$ factories, so that this method will
provide an important measurement of the charmless branching ratio.

Finally, an upper bound on $\BR(B \to X_{sg})$ can be obtained~\cite{CLEOBDDK} from 
the recent determination by CLEO of the ratio of ratios
\beq R  \equiv {\Gamma ( B \to \overline{D} X ) \over \Gamma(B \to all )} /
{\Gamma (B \to \overline{D} X \ell^+ \nu ) \over \Gamma (B \to X \ell^+ \nu ) } = .901 \pm .034 \pm .015,
\label{eq:Rratio} \eeq
using the relation
\beqa & R & =  1 + |V_{ub} /V_{cb} |^2 +  (\BR (B \to D_s^- \ell^+ X_\nu)  - \BR(B \to D_s^- X) \cr
&\!\! -\! \!&\!\! \BR (B\! \to\! (c \bar c) X )\! -\!\! \BR(B\! \to\! \Lambda_c^+ X)\! -\!\! \BR (B\! \to\! \Xi_c^{0,+} X)
\!-\!\!\BR (B\! \to\! X_{sg}) \eeqa
Unlike in the previous method, this bound does not depend on $\BR (D^0 \to K \pi )$.
Taking $.008 \pm .003 $ and $-.01 \pm .005 $ for the second and third terms~\cite{CLEOBDDK},
and using the most recent CLEO charmonium, $\Xi_{c}^{0,+}$, and flavor specific $\Lambda_c^+ $ yields,
which are given in Tables 1 and 2, we obtain
$\BR(B \to X_{sg}) = 1.7 \pm 4.4 \% $, or 
\beq \BR(B \to X_{sg}) < 9.0 \% ~~~~ @ 90\% c.l. \eeq
In Ref.~\cite{CLEOBDDK} older values for the charmonium and charmed baryon yields were used, 
leading to an upper bound of 6.8\%.  
From the measured lower ratio in $R$ a rather indirect determination of $\BR(D^0 \to  K^- \pi^+ )$
(they are inversely proportional) 
is possible, yielding~\cite{CLEOBDDK} $3.69 \pm 0.2\%$ which is lower than but consistent with the
world average for direct measurements. 
Motivated by the reasonable prejudice that the latter gives a more reliable
determination at this time, 
it is worth mentioning that a $1\sigma $ upwards shift in the 
lower ratio in $R$ would simultaneously reproduce the central value 
for direct measurements of $\BR(D^0 \to  K^- \pi^+ )$ and increase 
the central value for $\BR(B \to X_{sg})$ to 6\%.
It will be very interesting to see what 
this method ultimately yields for $\BR(B \to X_{sg})$ at the $B$ factories.

\subsection{Kaon counting}

It is possible to check whether the potentially large charmless yield is due 
to $b \to s$ transitions 
by comparing the measured flavor blind~\cite{pdg96} and flavor specific~\cite{ARGUSKaons} 
inclusive $\overline{B} \to K X$ branching ratios with 
the kaon yields from intermediate charmed states (see Refs.~\cite{kaoncharm,hawaii} for details).
The latter are divided into two classes:  
kaon yields which are essentially determined by
experiment and those which have to be estimated. 
For example, the largest known contributions are decays of intermediate $D/D_s$, 
which have been obtained by combining inclusive $\overline{B} \to DX /D_s X$ and PDG $D/ D_s  \to K X$ branching ratios.
Sizable $4.4 \sigma $ and $5.6 \sigma$ excesses remain in the total
$K^-$ and $K^+/ K^- $ yields, respectively, compared to known contributions.

The most important contribution to be estimated is $s \bar s $ popping in 
$\overline{B} \to X_{c \bar u d}$, leading to 
final states of the form $D K \overline{K} X$ and 
$D_s \overline{K} X$.\footnote{$s \bar s$ popping in other processes, e.g.,
$\overline{B} \to \Lambda_c X$
or $\overline{B} \to X_{c \bar c s}$,
can be safely neglected due to small rates for these processes
or phase space suppression.}  
The additional kaon yields per $\overline{B} \to X_{c \bar u d} \to DX /D_s X$
decay have been estimated
using a JETSET 7.4~\cite{jetset} string fragmentation model for $B \to X_{c \bar u
d}$. The total probability of $s \bar s $ popping in such decays
is found to be $\approx 14 \pm 3 \% $. 
Crude but generous estimates for kaon production from 
$\Lambda_c$, $\Xi_c$ and charmonium decays make up the rest.

Including the above estimates gives [\%]
\beqa \BR(\overline{B} \to K^- X) - \BR(\overline{B} \to X_c \to K^- X)
 &=  &16.8 \pm 5.6,~15.2 \pm 5.8 \nonumber\\
\BR(\overline{B} \to K^+ X) - \BR(\overline{B} \to X_c \to K^+ X)
& =  & 1.2 \pm 4.2,~0.8 \pm 4.2 \nonumber\\
\BR(\overline{B} \to K^+ /K^- X) - 
\BR(\overline{B} \to X_c \to K^+ /K^- X) & = & 17.8 \pm 5.3,~15.8 \pm 5.7
\nonumber \\
\BR(\overline{B}\!\! \to\!\! K^0 /\overline{K^0} X)\! - \!
\BR(\overline{B}\!\! \to\!\! X_c \to K^0 /\overline{K^0} X)\!\! & = &\!\! 4.6\pm 6.9, 
2.9 \pm 7.2 \label{eq:Kexcess} \eeqa
The second set of numbers is obtained using only the recent CLEO $B \to DX/D_s X$
branching ratios.
We see that a $3 \sigma $ $K^-$ excess remains. The $K^-$ excess is also 
reflected in the $3 \sigma $ total charged kaon excess.   
The $K^0/\overline{K^0}$ result is consistent with either no kaon excess, or sizable
kaon excess.  

The kaon excesses are consistent with
expectations from enhanced $b \to sg$.
Alternatively, if the 
excesses are due to underestimates of kaon yields    
then the most likely culprit is $s \bar s$ popping.
It is important to note that this can be determined
at the $B$ factories via measurements of
$\BR(\overline{B} \to D K \overline{K} X)$ and 
$\BR(\overline{B} \to D_s \overline{K} X)$. 
A measurement of the additional kaon spectra in these decays
is also extremely important since this is one of the largest 
uncertainties 
in determining the Standard Model inclusive kaon spectra.
About 90\% of the uncertainty in the charged kaon excesses in 
Eq.~\ref{eq:Kexcess} is due
to measurements of $\BR(B \to KX)$, $\BR(B \to D X)$, and $\BR(D \to KX)$. 
Fortunately, the first two branching ratios will be measured much more precisely at 
the $B$ factories.

\subsection{$n_c $ and $\BR(B \to X_{c \ell \nu_{\ell} })$}

\begin{figure}
\centerline{
\epsfbox{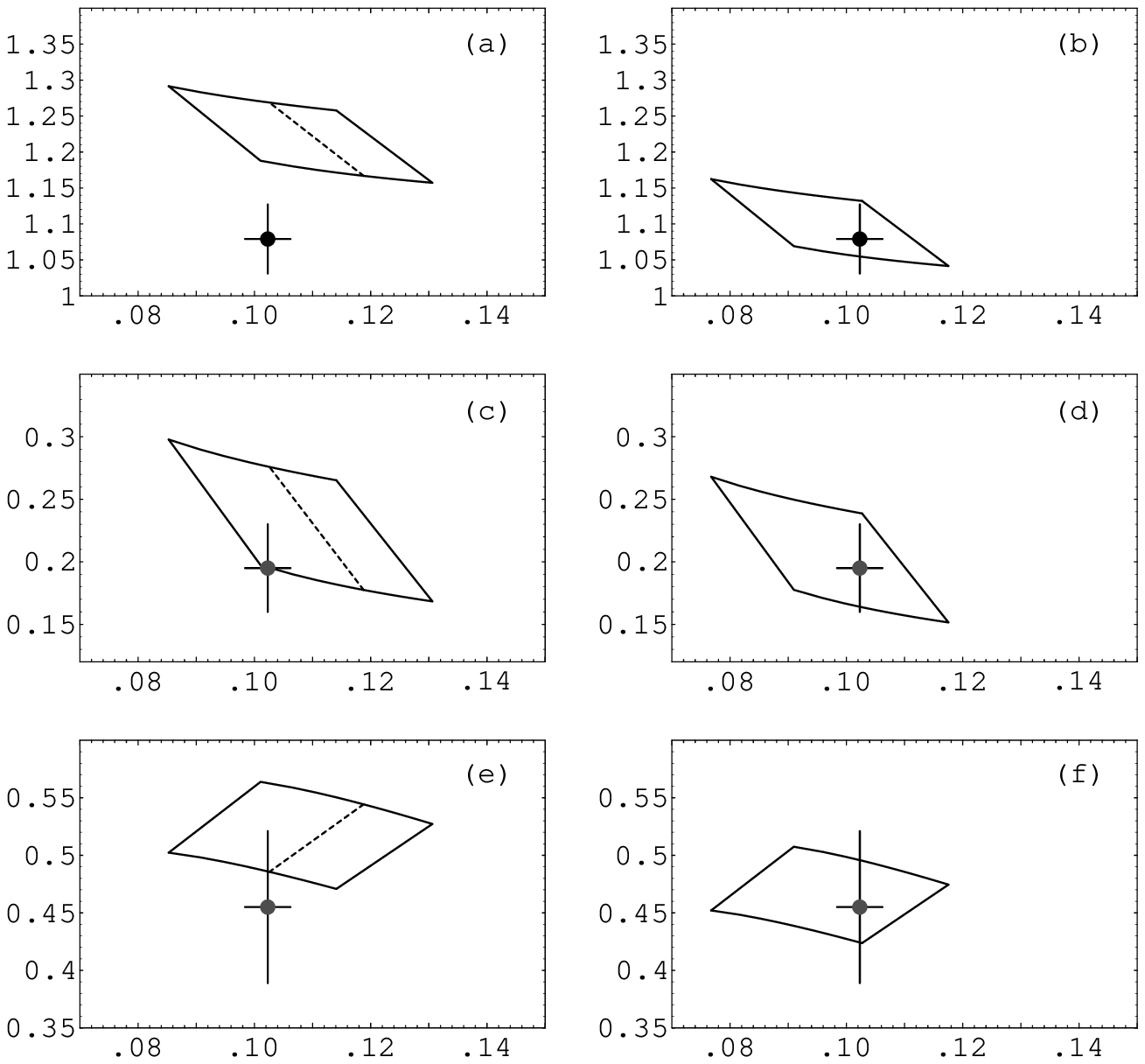}}
\caption{SM NLO predictions ($\alpha_s (M_Z) = .117$) for (a) $n_c$,
(c) $\BR(B\to X_{c \bar c s} )$, (e)  $\BR(B\to X_{c \bar u d} )$
vs. $\BR(B \to X_{c \ell \nu_{\ell} }) $. 
The impact of $\BR(b \to sg) = 10\% $ 
is shown in (b), (d), (f), respectively. 
Left (right) borders are for $\mu = m_b /4 $ ($\mu = m_b $). Dashed lines are for $\mu = m_b /2 $.
Bottom (top)
borders are for $m_b = 5.0 $, $m_c /m_b = .33$ ($m_b = 4.6$, $m_c /m_b = .25 $)
in (a) - (d). This is reversed in (e), (f). 
The crosses are the experimentally determined ranges at the $\Upsilon (4S) $.}
\label{fig:bsg1}
\end{figure}

In Fig.~\ref{fig:bsg1} predictions of the Standard Model and models with enhanced $b \to
sg$
for $n_c$, $\BR(\overline{B} \to X_{c \bar c s})$, 
$\BR(\overline{B} \to X_{c \bar u d})$, and $\BR(B \to X_{c \ell \nu_{\ell} })$
are compared with their measured values at the $\Upsilon (4S)$.  For the measured
semileptonic branching ratio we 
take~\cite{richman}
\beq \BR^{exp}(B \to X_{c \ell \nu_{\ell} }) = 10.23 \pm .39 , \label{eq:Bsl} \eeq
which is the average of the nearly model-independent 
ARGUS and CLEO dilepton charge
correlation measurements~\cite{bsl}.
The theoretical inputs include full on-shell scheme next-to-leading order QCD corrections~\cite{bagan}, 
and ${\cal O}(1/m_b^2)$ 
HQET corrections~\cite{bigi} to the tree-level $b \to c$ parton model decay
widths.
Next-to-leading order scheme-independent corrections to the $\Gamma (b \to c \bar c s)$ {\it penguin} 
contributions are also included.  The remaining
scheme-dependent corrections should be an order of magnitude smaller.
The $b \to u$ transitions have also been taken into account to NLO~\cite{bagan},
but the penguin $b \to s$ transitions have been neglected.
As in Ref.~\cite{neubertsachrajda} the quark pole masses are varied in the range   
$4.6 < m_b < 5.0$ and $.25 < m_c /m_b < .33$, and the renormalization scale is varied\footnote{The transition 
from 5 to 4 flavors is taken into account for $\mu < m_c$.}
from $m_b /4 < \mu < m_b$.

According to Figs.~\ref{fig:bsg1}a,c,e, $n^{exp}_c$ is lower than the 
Standard Model range for all $\mu$ but, as also pointed out in Ref.~\cite{neubertEPS}, 
$\BR^{exp} (B \to X_{c \bar c s})$
and $\BR^{exp}(B \to X_{c \bar u d})$ are consistent with 
the Standard Model ranges.  In particular, there is no indication
that the discrepancy in $n_c$ is due to poor theoretical control
over hadronic decays
beyond the sources of uncertainty already considered above, e.g., large deviations from local parton-hadron
duality.      
$\BR^{exp}(B \to X_{c \ell \nu_{\ell} })$ is consistent with the Standard Model range
at low values of $\mu$.  However, we caution that whereas 
a low renormalization scale appears to be justified for the semileptonic 
decay widths~\cite{beneke} this may not be the case for the hadronic decay widths
entering the branching ratio~\cite{hawaii}.

In Figs.~\ref{fig:bsg1}b,d,f we take $\BR(\overline{B} \to X_{sg})
\approx 
10\% $.  $n_c$ becomes consistent with experiment, 
$\BR^{exp} (B \to X_{c \bar c s})$
and $\BR^{exp}(B \to X_{c \bar u d})$ remain consistent with 
experiment, and $\BR^{exp}(B \to X_{c \ell \nu_{\ell} })$
can now be reproduced at larger values of $\mu$, i.e., without 
requiring large
perturbative or non-perturbative QCD corrections.

\section{$K$ production from $B \to X_{sg}$}

The main contribution of enhanced $b \to sg$ to kaon production is
fragmentation via soft
$q \bar q $ popping.  In the $b$ quark rest frame the gluon and $s$ quark emerge
back to back
with energy $m_b /2$.  In the string picture a string connects the $s$ and
spectator
quarks and the gluon is a kink in the string which carries energy and momentum~\cite{jetset}. 
The ensuing fragmentation is modeled~\cite{kaoncharm} 
using a JETSET 7.4 Monte Carlo with recent DELPHI tunings~\cite{hamacher}. 
The large energy released in $b \to sg$ decays should lead to
high multiplicity
final states, or soft kaon momentum spectra~\cite{swain}.  This expectation is confirmed by 
the Monte Carlo results, as we'll see below.
The number of kaons per $\overline{B} \to X_{sg} $ decay
produced in Monte-Carlo~\cite{kaoncharm} is .67 ($K^- $), .19 ($K^+ $), 
.62 ($\overline{K^0}$), and .15 ($K^0 $), so it is clear that 
enhanced $b \to sg$ would significantly reduce the charged kaon excesses.

Hard $q \bar q$ fragmentation of the gluon becomes important 
at large $K$ momenta.  In this case the 
decay $b \to s g* \to s q \bar q $ can be described by an effective four quark operator.  
The corresponding contribution to fast $K$ 
production can be estimated using factorization~\cite{deshhetram},
i.e., the meson is formed from the primary quarks in the decay. 
We return to the factorization model when discussing direct $CP$ violation.

\begin{figure}
\centerline{
\epsfbox{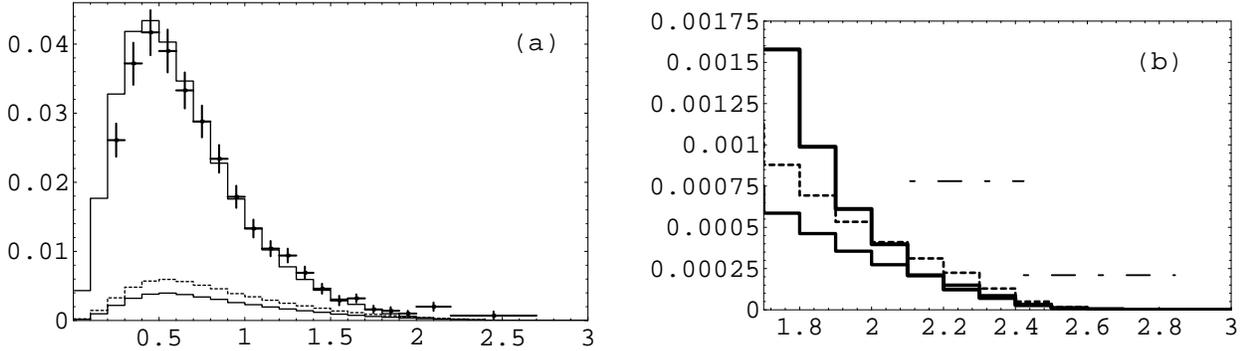}}
\caption{$\BR(B \to K_s X)$ vs. $p_{K_s}$ [GeV]. Branching ratios are 
for $0.1$ GeV bins except CLEO upper limits.  
(a) ARGUS data (crosses), 
SLD/CLEO Monte Carlo (upper solid), Monte Carlo for $\BR(\overline{B} \to X_{sg}) =
10\%$ 
(lower solid) and 15\% (dashed).  
(b) fast kaon spectra: CLEO 90\% CL UL{'}s for  $2.11 < p_{K_s} < 2.42$, 
$2.42 < p_{K_s} < 2.84$ (dot - dashed), 
SLD Monte Carlo (thick solid), Monte Carlo for $\BR(\overline{B} \to X_{sg}) = 10\%$ 
(solid), 15\% (dashed).}
\label{fig:bsg2}
\end{figure}

In Fig.~\ref{fig:bsg2} inclusive $K_s$ momentum spectra (in the $\Upsilon (4S)$ rest frame)
generated by
the $\overline{B}\to X_{sg}$ and
(SLD tuned) CLEO $\overline{B} \to X_c $ Monte Carlos are compared with
the measured spectrum~\cite{ARGUSKaons}. 
In the $b \to sg$ Monte Carlo the $b$ and spectator quark momenta are modeled
using the Gaussian distribtion of Ref.~\cite{altarelli}, with $p_F = 250~MeV$. 
Parton showers are also included.  
For those momenta where most $b \to sg$ kaons are produced the expected ratio of signal 
to Standard Model background is $\approx 1:5 - 1:10 $.  Clearly, 
resolving the presence of enhanced $b \to sg$ at these momenta {\it directly}
would be a very difficult task.  A vertexing veto of charm would
have to be extremely efficient to significantly 
enhance the $b \to sg$ component.
Perhaps the relative back-to-back geometry of signal events
versus the more spherical geometry of background events could help
discriminate between the two.  A related question is to what extent do
sphericity event shape cuts used to distinguish between continuum and $\Upsilon (4S)$ decays, 
such as those commonly employed by CLEO, bias against 
$b \to sg$ events.   

A promising strategy is to search for kaons from enhanced $b \to sg$
at higher momenta, e.g. $p_{K} \gsim 1.8~GeV$.  A 
significant kaon signal at still higher momenta, e.g., above 2.1 \gev, where the
background from intermediate charm states is highly suppressed  
would provide an unambigous signal for charmless $b \to s$ transitions~\cite{nussinov}.
Unfortunately, because of very large theoretical uncertainties above $2.1 ~GeV$
it would be very difficult to determine whether such a signal is due purely to Standard Model 
penguins, or intervention of enhanced $b \to sg$, unless
it happened to be close to the current upper bound~\cite{CLEOKsep}.  
According to Fig.~\ref{fig:bsg2} the ratio of enhanced $b \to sg$ signal 
to Standard Model background for $p_{K} \gsim 1.8~GeV$ is expected to be
$\sim 1:1$. Although branching ratios
are reduced to the $10^{-3}$ level, high statistics analysis will 
be possible at the $B$ factories.    
The background at large momenta
can be determined experimentally with little theoretical input. 
For example, the dominant $B \to D \to K$ contributions can be obtained
directly by folding measured $B \to D X$ and MARK III $D \to K X$  
inclusive momentum spectra. In fact, fast kaons mainly originate from
the lowest multiplicity $D$ decays\footnote{I thank Mark Convery for discussions 
of this point.}, e.g., $D \to K \pi,~K \rho,~K^* \pi$, which are well
measured, and the Cabbibo suppressed decays $B \to D K, D^* K$, which will be well measured in the 
future.
Of course, the $D$ spectra will be determined to very high precision at the $B$ factories. 
Furthermore, the poorly known $s \bar s$
popping and $B \to D_s \to K$ contributions should 
be much softer and are therefore unlikely to contribute significantly.
A vertexing veto of charm 
of modest efficiency, which should certainly be available at the $B$ factories, 
could significantly enhance the 
$b \to sg$ component above $1.8~GeV$.  It is worth mentioning that 
kaon momentum spectra obtained from a similar JETSET Monte Carlo for Standard Model
$b \to s \bar q q$ decays are similar in shape (again the large
energy release leads to high multiplicity decays or soft spectra).  However,
the parton level branching ratio is $\approx 1\%$ so
the kaon yields are an order of magnitude smaller than in the case of enhanced
$b \to sg$.

\begin{figure}
\centerline{
\epsfbox{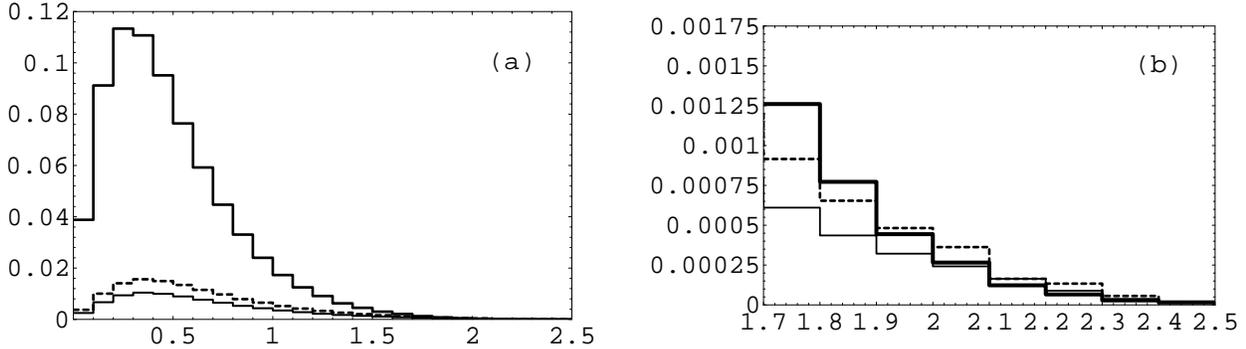}}
\caption {${\cal B}(B \to K^\pm X)$ vs. kaon $p_{T}$ [GeV]. 
Branching ratios are for $0.1$ GeV bins.  
(a) SLD/CLEO Monte Carlo (upper solid), Monte Carlo for ${\cal B}(\overline{B} \to X_{sg}) =
10\%$ 
(lower solid) and 15\% (dashed) with $p_F = 250$ MeV 
and showers included.  
(b) fast kaon spectra:  SLD Monte Carlo (thick solid), 
Monte Carlo for ${\cal B}(\overline{B} \to X_{sg}) = 10\%$ 
(solid) and 15\$ (dashed).}
\label{fig:bsg3}
\end{figure}

Fast kaon searches can be carried out at the $Z$ as well by studying
high $p_T$ $K^\pm$ production. 
According to Fig.~\ref{fig:bsg3} the expected signal to background ratio
for $p_T \ge 1.8~GeV$ is again $\sim 1:1$.
The SLD Collaboration has made 
a preliminary analysis using its '93-'95 and '96-'97 data samples~\cite{SLDeps}.
Unfortunately, the statistics are still too low to reach any definitive conclusions.  
Hopefully the situation will be further clarified at the 1998
summer conferences where a more extensive
analysis including more recent data will be presented.  
DELPHI has also searched for an excess of charged kaons in the $p_T$ spectrum~\cite{DELPHIeps} using
their high statistics samples.
However, to date DELPHI has pursued a different strategy, 
attempting to fit the entire measured spectrum 
with Monte Carlo $B \to X_{sg}$
and $B \to X_c $ components kept free. Unfortunately, this procedure suffers from 
too much model-dependence at the present time to reach definitive conclusions.
As is clear from Fig.~\ref{fig:bsg3}    
the shapes of the two components are not
expected to differ dramatically over a wide range of momenta, so 
they would have to be known 
fairly accurately in order to extract a $b \to sg$ contribution.

A significant excess relative to a well established
$B \to X_c \to K$ background rate at large kaon momenta would
provide model-independent evidence for new physics. 
The model-dependence would then enter when attempting to extrapolate the excess to 
lower momenta via the $B \to X_{sg}$ Monte Carlo, in order to determine the 
total charmless rate.
The Monte Carlo results for fast kaon production
can only be regarded 
as order of magnitude estimates.  
However, we should mention that JETSET succesfully describes the inclusive $K^\pm $ and $ K^0$ momentum spectra measured in 
$Z$ decays~\cite{ALEPHQCD}$^-$\cite{SLDQCD}, including the separately
measured spectra for $Z$ decays to light $(u,d,s)$ flavors \cite{DELPHIQCDEPS,SLDQCD}.
Unfortunately,
useful data at $x_p > .8$ ($x_p \equiv p_K /p_{beam} $) is not currently available.
JETSET also reproduces the measured portion of the 
$K^\pm$ spectrum in the $e^+ e^-$ continuum near the $\Upsilon$ 
resonances~\cite{ARGUSKaoncont}, however in this case there is no data available at 
$x_p  > .5 $.  
Precision high $x_p$ data will be very important
for further tuning of Monte Carlo fragmentation parameters relevant
to {\it fast kaon} production in $b \to s $ decays, and should
be available from $B$ factory continuum studies.  
Spectra for kaons
produced in the continuum at a charm factory, e.g., $\sqrt{s} \sim 4~GeV$, 
would of course be extremely useful.  We also note that
the $B \to X_{sg}$ Monte Carlo fast kaon yields obtained with 
default JETSET tunings are approximately 30\% smaller than those obtained with 
the recent DELPHI tunings~\cite{hamacher}, which gives an idea of the large systematic
errors involved.  Finally, fast kaon production is
extremely sensitive to the modelling of Fermi motion. For example, setting 
the Fermi momentum parameter $p_F = 250~MeV$
in the Gaussian model of Ref.~\cite{altarelli} reduces production of kaons with momenta
above $1.8~GeV$ by about 50\% compared to production without Fermi motion.
More sophisticated treatments of Fermi motion are therefore required.

\section{$CP$ violation}

Enhanced dipole operator coefficients can carry new $CP$ violating
weak phases.  
Furthermore, for $\BR(b \to sg) \sim 10  \% $ 
the dipole amplitudes for rare hadronic decays are of same order
as the Standard Model amplitudes~\cite{hawaii}.  
Since the strong interaction phases associated with the two amplitudes must in general differ, 
their interference can lead to large direct $CP$ asymmetries ($A_{CP}^{dir}$).
We present factorization model results for 
$B^\pm \to \phi K^\pm$, $B \to \phi X_s $, $B^\pm \to K^0 \pi^\pm$ and
$B^0 \to K^\pm \pi^\mp $.  
For the moment we will ignore soft final state interactions (FSI), which
actually are expected to be important in $B$ 
decays~\cite{donoghue}$^-$\cite{soniFSI}.
In the absence of FSI 
the Standard Model contributions for the first three modes above are dominated by the
penguin $b \to s \bar s s$ and $b \to s \bar d d$ transitions, resulting in
small $CP$ violating asymmetries~\cite{fleischer} of order $1 \%$.

The relevant $\Delta B =1$ effective weak Hamiltonian takes the form~\cite{burasQi}
\beq {\cal H}_{\rm eff}  = \frac{G_F}{\sqrt{2}}\left[
 V_{ub} V_{us}^*  \sum_{i=1}^{2}
c_i Q_i^{u} - V_{tb} V_{ts}^* \left(\sum_{i = 3}^{11} c_i Q_i + c_{11}' Q_{11}' \right) 
\right].
\label{Heffweak}\eeq
The flavor structures of the current-current, QCD penguin, and electroweak
penguin operators are, respectively,
$Q_{1,2}^{u} \sim \bar s u  \bar u b $,
$Q_{3,..,6} \sim \bar s b \sum \bar q' q' $, and
$Q_{7,..,10} \sim \bar s b \sum e_{q'} \bar q' q' $, where the sum is over
light quark flavors.  The chromomagnetic dipole operators are given by
\beq Q_{11} = \frac{g_s}{16 \pi^2} m_b (\mu)
\bar s \sigma_{\mu \nu} R t^a b G^{\mu \nu}_a~, ~~ Q_{11}' = \frac{g_s}{16 \pi^2}
m_b (\mu) \bar s \sigma_{\mu \nu} L t^a b G^{\mu \nu}_a . \eeq
They are
included in the factorization model by allowing 
the gluon to go off shell and turn into a quark-antiquark pair~\cite{deshhetram,hayashi}.  
We parametrize the dipole operator Wilson coefficients as
\beq c_{11} = - |c_{11} | e^{i \theta_{11} },~~~~
c'_{11} = - |c{'}_{11} | e^{i \theta{'}_{11} }. \label{eq:theta11} \eeq
In the standard model $c_{11} (m_b) \approx - .15$, and
$c{'} _{11} $ is a factor $m_s /m_b$ smaller, hence negligible. $\BR(b \to sg)
\sim 10\%$ requires $(|c_{11} |^2 + |c'_{11} |^2)^{1\over2}$
to be enhanced by nearly an order of magnitude.  
Strong phases originating at NLO from $c \bar c$ rescattering~\cite{bander}
have been taken into account in the Standard Model penguin amplitudes using the 
NLO scheme-independent effective Wilson coefficient 
formalism of Refs.~\cite{fleischer,deshhe}. 
For the numerical inputs we choose $\alpha_s (m_b) = .212$,
$m_b = 4.8~GeV$, $m_c = 1.4~GeV$, $m_s (m_b) = 0.1~GeV$, 
and $N^{eff}_c = 10$ for the effective
number of colors parametrizing non-factorizable corrections (called $1/\xi$ in Ref.~\cite{BSW}).  
We also take $\rho = .11$ and
$\eta = .33$ for the Wolfenstein parameters, 
$\mu = m_b$ for the renormalization scale, 
and $q^2 \approx m_b^2 /2$ for the square of the virtual gluon momentum
entering the penguin and dipole amplitudes. For the $B \to K, \pi $ form factors
at zero momentum transfer    
we take $F_{1,0}^K (0) = F_{1,0}^\pi (0) = .33 $ in the standard parametrization of Ref.~\cite{BSW}.
A dipole behaviour is taken for the $q^2$ dependence of $F_1 $ and a monopole 
behaviour is taken for $F_0$, according to the lattice results given in Ref.~\cite{lellouch}.

\begin{figure}
\centerline{
\epsfbox{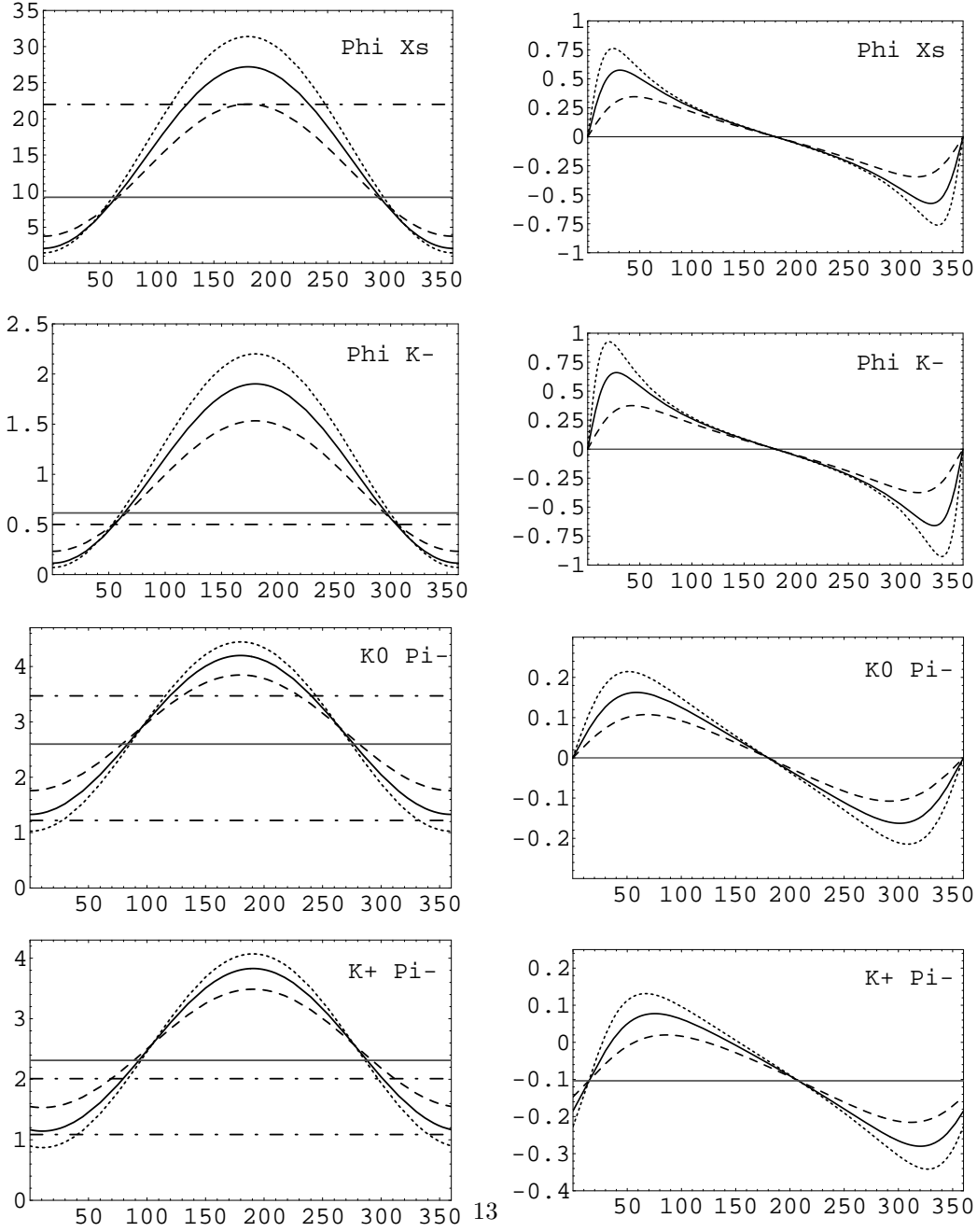}}
\caption{$10^5 \BR(B \to \phi X_s,\phi K^\pm,K^0 \pi^\pm, K^\pm \pi^\mp)$
($CP$ averaged) vs. $\theta_{11}$ 
and corresponding direct $CP$ asymmetries vs. $\theta_{11}$ for 
$\BR(b \to sg) \approx 15\%$ (dotted curves), 10\% (solid curves), 5\% (dashed curves).
Horizontal solid lines are the corresponding Standard Model branching ratios and asymmetries. 
Horizontal dot-dashed lines 
are CLEO 90\% c.l. branching ratio upper limits for $B \to \phi K^\pm, \phi X_s $
and measured $1 \sigma$ ranges for $B \to K^0 \pi^\pm, K^\pm \pi^\mp $.}
\label{fig:bsg4}
\end{figure}

\begin{figure}
\centerline{
\epsfbox{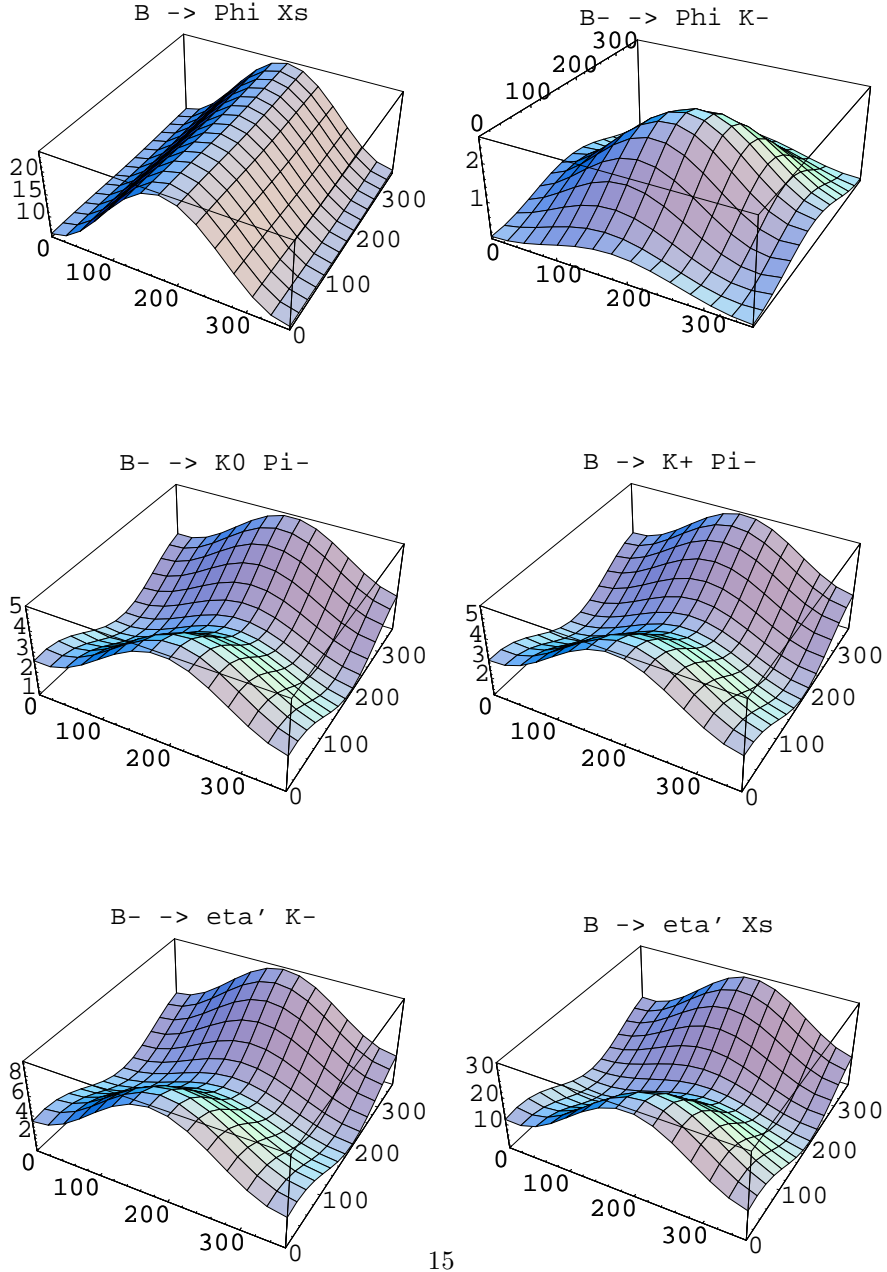}}
\caption{
$10^5 \BR(B \to \phi X_s,\phi K^\pm,K^0 \pi^\pm, K^\pm \pi^\mp)$
 vs. $\theta_{11}$ (left axes) and  $\theta_{11}'$,
for $|c_{11}'| = |c_{11}|$ and $\BR(b \to sg) \approx 10\%$.
Also included are $ 10^5 \BR(B^\pm \to \eta' K^\pm )$ and the quasi two-body
contribution to $10^5 \BR(B \to \eta' X_s)$ in the factorization model, as discussed
in Section 5. All branching ratios are $CP$ conjugate averages.}
\label{fig:bsg5}
\end{figure}

In Fig.~\ref{fig:bsg4} branching ratios ($CP$ averaged) 
and direct $CP$ asymmetries obtained with enhanced 
$c_{11}$ and $c_{11}' = 0$ are compared to
Standard Model predictions. Also included are CLEO measurements
or upper bounds for the branching ratios~\cite{CLEOphiep}$^-$\cite{smith}. 
In the case of $B \to \phi X_s $ the bound has been obtained for
$p_\phi > 2.1~GeV$.  
The large dependence on $\theta_{11}$ confirms that there can be
substantial destructive or constructive interference between 
the penguin and dipole amplitudes.  
Given the large uncertainties in the factorization
model estimates, i.e., sensitivity to numerical inputs, 
non-factorizable contributions (parametrized by $N_c$), and absence of FSI,
it is clear that the experimental branching ratio constraints 
can be satisfied for a significant range of $\theta_{11}$, even if
$\BR(b \to sg) \approx 15\%$.
Although the charmless branching ratio with enhanced $b \to sg$ is an order of magnitude larger 
than in the Standard Model it is peaked at very low $q^2$, i.e., the 
``on-shell" gluon limit.  At larger values of $q^2$ relevant to two-body
or quasi two-body rare decays the enhanced $b \to sg$ amplitudes are
reduced to the level of the Standard Model amplitudes, as is well illustrated by a 
parton level Dalitz plot analysis of $b \to s \bar q q$ decays~\cite{hawaii,kaganpetrov}.   
We should mention that in the Standard Model the
interference of the dipole and penguin amplitudes is destructive  
for the charmless $b \to s \bar q q$ transitions at the parton level~\cite{nierste}, 
leading to a reduction in all corresponding 
factorization model decay rates by $\sim 10-20\% $.
In general in models 
with enhanced $b \to sg $ there is no reason to expect $c_{11} >> c_{11}' $
in the absence of extra flavor symmetries.  
As an illustrative example, in Fig.~\ref{fig:bsg5} we show the 
branching ratio dependences on $\theta_{11},~\theta_{11}' $ for 
$ |c_{11}| = |c_{11}'| $ and $\BR(b \to sg) \approx 10\% $.  Thus, we see that
the general case where both $Q_{11}$ and $Q_{11}'$ are enhanced is even less constrained.

It is clear from Fig.~\ref{fig:bsg4} that in the case of 
the semi-inclusive $\phi X_s$ and exclusive
$\phi K^\pm, K^0 \pi^\pm$ modes enhanced $b\to sg$ can lead to 
sizable direct $CP$ asymmetries\footnote{In Ref.~\cite{hawaii} 
the CP asymmetries were incorrectly given with opposite sign due to 
a sign error in the $c \bar c$ rescattering strong phases.}
which are easily more than an order of magnitude
larger than the naive ${\cal O}(1\%)$ asymmetries expected in the standard model 
when ignoring soft FSI. Large contributions to the $B \to \phi K_s$
time dependent $CP$ asymmetry
would also arise~\cite{worah}$^-$\cite{valencia}.
However, as has been discussed recently~\cite{gerard}$^-$\cite{soniFSI}, 
the impact of FSI on $A_{CP}^{dir} (B \to K \pi )$ 
is likely to be quite sizable.  For example, in Ref.~\cite{falkFSI} a crude 
model based on Regge phenomenology
was considered in which only two-body pseudoscalar intermediate states were taken into account.
Rescattering contributions of the 
doubly Cabbibo suppressed $b \to u \bar u s$ current-current operators 
to the $B^+ \to K^0 \pi^+ $
amplitude were found to be of order 10\%. Furthermore, 
since almost all inelastic contributions were neglected the
actual contributions could be significantly larger.
It is therefore not inconceivable that in the Standard Model 
$A_{CP}^{dir} (B^+ \to K^0 \pi^+ )$ could be several tens of percent.
It stands to reason that similar rescattering 
contributions to the $B^+ \to \phi K^+ $ amplitude 
could also be of order 10\%, again 
leading to sizable
asymmetries.

Clearly, large direct CP violation in the exclusive modes $B^\pm \to \phi K^{\pm},
K^0 \pi^\pm$ would not provide an unambigouos signal for New Physics.
Fortunately, it is possible to use flavor $SU(3)$ tests to bound the magnitudes of FSI
contributions and Standard Model asymmetries 
in these modes along the lines suggested in~\cite{grossmanisidori} for the $B \to \phi K_s$ time-dependent asymmetry.
A bound which could turn out to be particularly useful since it 
only depends on the ratio  
\beq\label{RK} R_K = {\BR(B^+ \to K^+ \overline{K^0} ) +
\BR(B^- \to K^- K^0 ) \over \BR(B^+ \to K^0
\pi^+ ) +\BR(B^- \to \overline{K^0}  \pi^- )}\, \eeq
is~\cite{falkFSI} 
\beq |A_{CP}^{dir} (B^\pm \to K^0 \pi^\pm )| < 2 \lambda \sqrt{R_{K}}\, 
(1 + R_{SU(3)} )
+ {\cal O}(\lambda^3 , \lambda^2 R_{SU(3)} )\,. \eeq
$\lambda \cong .22$ is the Wolfenstein parameter, and $R_{SU(3)}$ 
parametrizes $SU(3)$ violation and is typically of order 20\% - 30\%.
The analogous bound for $A_{CP}^{dir} (B^\pm \to \phi K^\pm )$ follows
by substituting for $R_K$ the ratio 
$[\sqrt{\BR(B^\pm \to K^{0(*)} K^\pm )} + 
\sqrt{\BR(B^\pm \to \phi \pi^\pm)} ]^2 /\BR(B^\pm \to \phi K^\pm )$.
Direct CP asymmetries in excess of these bounds would provide a clear 
signal for New Physics.  It should be noted that analogous tests for New Physics 
are possible in other modes, e.g., $B^\pm \to  \phi K^{*\pm}, K^{0 *} \pi^\pm$.
Alternative startegies employing
$B \to K \overline{K} $ decays
to constrain FSI contributions to $B \to K \pi$ are discussed in 
Ref.~\cite{fleischerFSI}.

\section{$B^\pm \to \eta' K^\pm$ and $B \to \eta' X_s $}

The CLEO collaboration has reported an exclusive rate~\cite{cleoetaprK,smith}
\begin{equation} 
{\cal B}(B^{\pm} \to \eta'  K^{\pm}) = (6.5^{+1.5}_{-1.4} \pm 0.9)
\times 10^{-5}, 
\label{eq:cleoexc}\end{equation} 
and a semi-inclusive rate~\cite{CLEOetaprX}
\begin{equation} 
{\cal{B}}(B \to \eta'  X_s ) =  (6.2 \pm 1.6 \pm 1.3 ^{+0.0}_{1.5}) \times 10^{-4} 
~(2.0 < p_{\eta'} < 2.7~GeV). 
\label{eq:cleoinc}\end{equation}
The experimental cut on 
$p_{\eta'}$ is beyond the
kinematic limit for most $b \to c$ decays 
and corresponds to a recoil mass $m_{X_s} <2.5~GeV$ in the laboratory frame. 
There is no evidence for $\eta' K^*$ modes in the inclusive analysis, with most
events lying at large recoil mass\footnote{The signal yield consists of
$11 \pm 8.4$ events for $m_{X_s} < 1.8$ GeV, and $27.5 \pm 7.8$ events
for $1.8 < m_{X_s} < 2.5$ GeV.}, $m_{X_s} > 1.8~GeV$.  
The surprisingly large branching ratios have prompted many phenomenological papers
on $\eta'$ production, 
some of which are 
listed here~\cite{aligreub}$^-$\cite{kao,kaganpetrov}.
A consensus has emerged 
that the exclusive branching ratio measurement can
be accounted for in the Standard Model in the factorization 
approach~\cite{aligreub,deshduttaoh,chengtseng2,dattahepakvasa,kaganpetrov}.
However, the inclusive branching ratio is more problematic.
We will discuss the impact of enhanced $b \to sg$ on both rates in the factorization
model.  In the inclusive case we will also discuss the gluon anomaly 
mediated subprocess $b \to sg^* \to \eta' s g$, first investigated in Ref.~\cite{atwoodsoni},
and potential long-distance contributions.  The results 
presented here are part of a collaboration with Alexey Petrov~\cite{kaganpetrov}.

\subsection{$B^\pm  \to \eta' K^\pm$: Factorization Model Contributions}

We adopt the two-angle mixing formalism for
$\eta - \eta'$ mixing~\cite{leutwyler,feldmannkroll}, 
\beq |\eta \rangle = cos\theta_8 |\eta_8 \rangle  - sin \theta_0 
|\eta_0 \rangle,~~~|\eta' \rangle = sin\theta_8 |\eta_8 \rangle  + cos\theta_0 
|\eta_0 \rangle~.\eeq
Our numerical inputs for the mixing angles and $SU(3)$ octet and singlet axial 
vector current decay constants
are the best-fit values found in~\cite{feldmannkroll}:
\beq  \theta_8 = -22.2^o,~~\theta_0 = -9.1^o,~~{f_8 \over f_\pi} = 1.28,~~
{f_0 \over f_\pi} = 1.20 . \eeq
For simplicity, we set $\langle \eta ' |\bar c \gamma_\mu \gamma_5 c |0\rangle = 0$ 
as it has become clearer~\cite{feldmannkroll}$^-$\cite{petrov,aligreub,chengtseng2}
that the intrinsic charm content 
of the $\eta'$ is not large enough to play a crucial role in $\eta' $ production.
The anomaly is taken into account in
the matrix element~\cite{tytgat,kaganpetrov} $\langle \eta' |\bar s \gamma_5 s |0\rangle$.
$SU(3)$ symmetry is imposed to relate the $B \to \eta'$ form factors 
\footnote{In the original version of Ref.~\cite{kaganpetrov} the
normalization constant of the $\eta'$ wave function was not taken into account in the form factors 
$F_{0,1}^{\eta'}$.}    
$F_{0,1}^{\eta'}$ at zero momentum transfer to $F_{0}^\pi (0)$, 
as in~\cite{aligreub,chengtseng1,deshduttaoh,chengtseng2}.  For the $q^2 $ dependences 
of $F_{0,1}^{\eta'}$ we again use the monopole and dipole
behaviours given in Ref.~\cite{lellouch}.  
Other inputs entering our factorization 
model calculations have been discussed in Section 4.  
  
\begin{figure}
\centerline{
\epsfbox{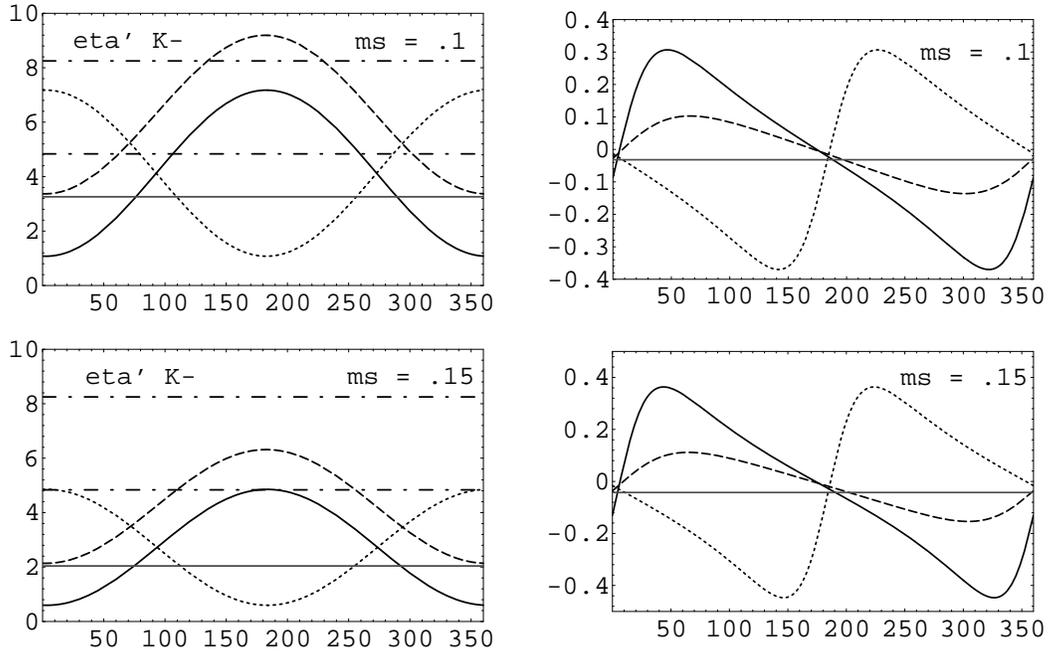}}
\caption{
$10^5 \BR(B^\pm \to \eta' K^\pm)$ ($CP$ averaged) and 
$A_{CP}^{dir} (B^\pm \to \eta' K^\pm)$ 
 vs. $\theta$ for $\BR(b \to sg) \approx 10\%$ and
 $m_s = 100~MeV$, $150~MeV$.  Solid curves are for $c_{11}' = 0$ 
($\theta = \theta_{11}$), 
long dashed curves are for  $|c_{11}| = |c_{11}'|$ and 
$\theta_{11}' = 180^o$ ($\theta = \theta_{11}$), short 
dashed curves are for $c_{11} = 0$ ($\theta = \theta_{11}'$). }
\label{fig:bsg6}
\end{figure}

In Fig.~\ref{fig:bsg6} we compare $\BR(B^\pm \to \eta' K^\pm)$ ($CP$ averaged)
and $A_{CP}^{dir} (B^\pm \to \eta' K^\pm ) $ 
obtained with $\BR(b \to sg) \approx 10\%$ to
the corresponding Standard Model values in the factorization
model. The observed $\pm 1 \sigma$ range for the branching ratio is also included.
Also see Fig.~\ref{fig:bsg5} for a plot of the branching ratio in the $(\theta_{11}, \theta'_{11})$ plane, given
$|c_{11}| = |c_{11}'|$.
One feature that stands out immediately is the sensitivity
of the exclusive rate to the current mass $m_s $, 
first noted in Ref.~\cite{kaganpetrov}.
This is because both $\langle \eta' | \bar s \gamma_5 s |0 \rangle$
and $\langle K^- | \bar s \gamma_5 u |0 \rangle$ are inversely 
proportional to $m_s$ via the equations of motion. 
Recent lattice and QCD sum rule studies give~\cite{msvalues1}
$\overline{m_s} (2~GeV) \approx 128\pm 18~MeV$ and~\cite{msvalues2}
$\overline{m_s} (2~GeV) \approx 100\pm 21 \pm 10~MeV$,
so that $m_s (m_b) \sim 100~MeV$ appears to be a reasonable choice.
Given the large uncertainties inherent in the factorization approach
it appears that the observed rate, particularly the lower 
half of the quoted range, can be accomodated in the 
Standard Model.  For example, it has recently been shown~\cite{chengtseng2}
that with a more general parametrization of non-factorizable corrections in which
two different parameters $N^{eff}_c (V-A)$ and $N^{eff}_c (V+A)$ are assigned to matrix elements
of four quark operators with $(V-A) (V-A)$ and $(V-A)(V+A)$ structure, respectively,
it is possible
to increase the Standard Model rate by more than 50\% while satisfying constraints on 
other exclusive decays.
However, it is also clear that enhanced $b \to sg$ can significantly 
increase the rate.  
$A_{CP}^{dir}$ can receive large contributions from enhanced $b \to sg$ 
thus leading to 
asymmetries which are significantly larger 
than the naive Standard Model factorization estimates of a 
few percent~\cite{deshduttaoh,petrov}.
However, one should not take Standard Model asymmetry estimates obtained in the factorization
approach too seriously, particularly because of the unknown impact of soft FSI.

\subsection{ $B\to \eta' X_s $: Factorization Model Contributions}

Factorization model contributions to $\overline{B} \to \eta' X_s $ involve two
types of amplitudes distinguished by their hadronization pattern, namely
quasi two-body and quasi three-body decays. 
In the two-body decays an $\eta' $ is formed from $s \bar s$, $u\bar u $
or $d \bar d$ pairs 
via the subprocess $b \to s q \bar q$, which in the parton model corresponds to
$b \to \eta' s$.  
The three-body decays involve hadronization of the spectator into the $\eta'$, 
which in the parton model
corresponds to $\overline{B} \to \eta' s \bar q$ ($q = u,d$). 
The two types of decay give rise to very different recoil spectra. 
In two-body decays $E_{\eta'} \sim m_b/2 $ and $q^2 \sim m_b^2 /2 $,
leading to 
a recoil spectrum peaked at low energies, e.g., 
$m_{X_s} \sim 1.4~GeV$, with a spread 
of a few hundred $MeV$ when Fermi motion is taken into account. 
In three-body decays the recoil spectrum rises steadily 
with $m_{X_s}$ and actually peaks beyond the signal region~\cite{kaganpetrov}.
\footnote{A Dalitz plot analysis
of quark level $b \to s \bar q q$ decays~\cite{kaganpetrov,hawaii} 
suggests that interference 
between the two-body and three-body amplitudes can be neglected.
In any case this interference would have negligible impact at large $m_{X_s}$
where most of the signal is located.}

\begin{figure}
\centerline{
\epsfbox{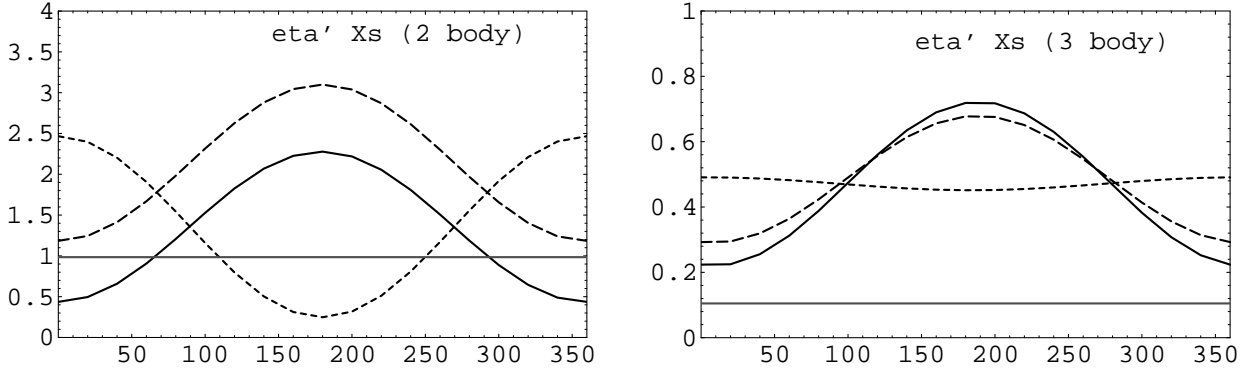}}
\caption{
$10^5 \BR(b \to \eta' s)$ and $10^5 \BR(\overline{B} \to \eta' s \bar q)$
($CP$ averaged) 
 vs. $\theta$ for $\BR(b \to sg) \approx 10\%$ and
 $m_s = 100~MeV$.  Solid curves are for $c_{11}' = 0$ 
($\theta = \theta_{11}$), 
long dashed curves are for  $|c_{11}| = |c_{11}'|$ and 
$\theta_{11}' = 180^o$ ($\theta = \theta_{11}$), short 
dashed curves are for $c_{11} = 0$ ($\theta = \theta_{11}'$).
Horizontal lines are Standard Model branching ratios.}
\label{fig:bsg7}
\end{figure}

In Fig.~\ref{fig:bsg7} two-body and three-body 
branching ratios with $\BR(b \to sg) \approx 10\%$ are compared 
to the corresponding Standard Model 
branching ratios for the inputs discussed above.\footnote{The
$\sim \alpha_s^2  c^2_{11} $ contributions
to $\Gamma (\overline{B} \to \eta' s \bar q) $ and 
$\Gamma (b \to s \bar q q )$ diverge in the $q^2 \to 0$ ``on-shell" gluon limit.
At the quark level 
this divergence must be canceled by the 
${\cal O}(\alpha_s) $ correction to $\Gamma (b \to sg )$.
The three-body $\eta '$ yield has been obtained with a cutoff
$q^2 > 1~{\rm GeV}^2$ imposed on the virtual gluon momentum.  The dependence 
on cutoff is negligible in the Standard Model, and is
sufficiently moderate near $1~{\rm GeV}^2$ in the case of enhanced $b \to sg$, i.e.,
the branching ratio decreases by 25\% as the cutoff is increased form 
1 ${\rm GeV}^2$ to 2 ${\rm GeV}^2$.}
A plot of the two-body branching ratio in the 
$(\theta_{11}, \theta'_{11})$ plane with
$|c_{11}| = |c_{11}'|$ is also included in  Fig.~\ref{fig:bsg5}.
For the three-body branching ratios 
we have for simplicity taken  $p_{\eta'} > 2~{\rm GeV}$ 
in the $B$ rest frame, thus ignoring the small effect
of the Lorentz boost to the laboratory frame. 
In the Standard Model we obtain $\BR(b \to \eta' s ) \approx 1 \times 10^{-4}$
and $\BR(B \to \eta' s \bar q) \approx 1.4 \times 10^{-5}$.  
A similar result was
obtained in Ref.~\cite{dattahepakvasa}.  
In models with enhanced $b \to sg$ both yields can be substantially increased.  
However, because
the two-body contribution is generally larger
than the three-body contribution by a factor 2 to 4
we conclude on the basis of the observed recoil
spectrum~\cite{CLEOetaprX}
that the bulk of the signal 
can not be accounted for in the factorization approach in models
with enhanced $b \to sg$.

\subsection{$B \to \eta' X_s $: Anomaly Mediated Processes}

It has been suggested by Atwood and Soni~\cite{atwoodsoni} (AS)  
that the large inclusive rate 
is connected to the QCD penguins via the gluon anomaly,
leading to the subprocess $b \to sg^* \to \eta' s g$.
The effective $gg\eta'$ coupling was parametrized as
\begin{equation}
V_{\mu \nu} \epsilon_1^\mu \epsilon_2^\nu =
H(q^2, k^2, q_{\eta^\prime}^2) \epsilon_{\alpha \beta \mu \nu}
q^\alpha k^\beta \epsilon_1^\mu \epsilon_2^\nu
\end{equation}
where $q^2 \equiv (p_b-p_s)^2$.
A constant form 
factor was assumed,
i.e.,
$H(q^2, k^2, q_{\eta'}^2) \simeq H(0, 0, m_{\eta'}^2) 
\equiv H_0$, and $H_0$ was extracted directly 
from the decay rate for $J/\psi \to \eta' \gamma$,
yielding $H_0 \approx 1.8 ~GeV^{-1}$.
With the central assumption of a constant form factor, 
AS found $\BR(b \to \eta' s g) \sim 8\times 10^{-4}$ ($p_{\eta'} > 2~GeV$), 
which would account for the observed signal.
Moreover, the three-body decay leads to an $m_{X_s}$ spectrum that is consistent
with observation. 
However, it is clear that in order to obtain the total decay rate
the differential distribution for this subprocess 
must be integrated over a wide range of
$q^2$, spanning approximately 1 $GeV^2$
to $m_b^2$. It is therefore of paramount importance to investigate 
the off-shell $q^2$ dependence of the form-factor.

Hou and Tseng~\cite{houtseng} (HT) argued that the factor $\alpha_s$
implicit in $H$ should be running and must therefore be 
evaluated at the scale of momentum transfer through the $\eta' g g$
vertex.  The mild logarithmic dependence of $H$ on $q^2$ 
would lower AS's result by roughly a factor of 3. 
HT therefore suggested that enhanced $b \to sg$ might be required in order to account
for the observed $\eta'$ signal, noting that this could substantially increase 
the $b \to s g \eta'$ rate.  They have also made the
interesting observation that this could lead to direct CP asymmetries for this 
subprocess which are as large as $10\%$.
Still, given such mild $q^2$ dependence we note that 
a Standard Model cocktail solution
could account for the observed signal, 
particularly when the large hadronic uncertainties 
involved are considered.  The introduction of New Physics is therefore not warranted in this
case.  
For example, the combination of 
a decreased $b \to s g \eta' $ branching ratio of $(2 - 3) \times 10^{-4}$, 
a factorization model branching ratio of $1\times 10^{-4}$, 
and a branching ratio of $1.1 \times 10^{-4}$ (with experimental cut)
for $\eta'$ production via decay of intermediate charmonia~\cite{atwoodsoni}
would be consistent with the observed $\eta'$ yield 
at the $1 \sigma$ level.  Furthermore, since 
$b \to \eta' s g$ and intermediate charmonia decays
lead to recoil spectra whose shapes are consistent with observation the total recoil
spectrum could also be consistent.

In Ref.~\cite{kaganpetrov} we suggested that the leading $q^2 $ dependence of the $g g \eta'$
form factor is much more severe
than assumed in~\cite{atwoodsoni,houtseng}.
First, we observe that the ``direct'' 
$gg \eta'$ coupling is suppressed.  Since the quantum 
numbers of the $\eta'$ and gluon are $0^-$ and $1^-$, respectively, 
formation of an $\eta'$
requires orbital angular momentum.  This introduces a
dependence on the transverse gluon momentum, so that
asymptotically the coupling 
scales like~\cite{baiergrozin,feldmannkroll} $1/q^4 $.
The leading order contributions in $1/q^2$ 
are, in fact, due to hard 
amplitudes involving quark or gluon exchange which
couple to the $|\bar q q \rangle $ or $|gg \rangle $ components of the
$\eta'$, respectively.
To model the leading $q^2$ dependence in the region of interest we
therefore consider a $gg \eta'$ vertex in which a pseudoscalar current is coupled 
perturbatively to two gluons through
quark loops.  

The calculation
yields a form factor which can be parametrized as
\beq H (q^2,0,m_{\eta'}^2) = -\frac{H_0  m_{\eta'}^2}{q^2 -m_{\eta'}^2}.
\label{eq:formfactor}
\eeq
In the general case of two virtual
gluons with momenta $q$ and $k$ the denominator is replaced with $2 q \cdot k $. 
Recently, it has been shown that quark condensate contributions
do not modify the leading $q^2$ dependence, suggesting that it 
is not modified by non-perturbative QCD effects~\cite{ahmady}.
The dependence of $H_0$ on $q^2$, which includes 
possible running of $\alpha_s$, must be subleading compared to the 
strong power dependence. However, it 
insures the absence of a pole at $q^2 = m_{\eta'}^2 $. 
To first approximation it can be modeled by a constant $H_0$ 
which we identify with the value extracted from $J/\psi \to \eta' \gamma$. 
In the Standard Model this leads to 
$\BR( b \to s g \eta' ) \sim 1.6 \times 10^{-5} $ including cuts (for $m_b = 5.0~GeV$),
which is more than an order of magnitude smaller than the 
observed $\eta' $ yield.  On the other hand, for 
$\BR(b \to sg) \approx 10\% $ we obtain  
$\BR( b \to s g \eta' ) \sim .2 - 1.5 \times 10^{-4} $, with the largest 
values, as usual, corresponding to $\theta_{11} \sim 180^o $.
The corresponding direct CP asymmetries
are systematically smaller than those obtained by HT because of the shift in
the decay distribution to lower $q^2$, thus leading to smaller strong phases 
in the chromoelectric form factor.

As a further indication that $H(q^2 , 0 , m_{\eta '}^2 )$ has a 
strong power dependence we note
that precisely the same  $1/(q^2 -m_{\eta'}^2)$ behaviour is obtained 
via quark exchange in the non-relativistic quark model~\cite{kuhn,close}.
Furthermore, the leading $q^2$ dependence associated with gluon exchange
(or a gluon ``loop")
will not be any less severe, as can be seen from the form of the gluon propagator.
HT have 
suggested\footnote{See note added in Ref.~\cite{houtseng}.} that 
intermediate 
gluonia arising via the gluon self coupling 
might postpone the onset of $q^2 $ suppression by giving rise to a 
$1/(q^2 - m_G^2 )$ dependence in the form factor, 
where $m_G$ is a glueball mass scale of $2-4~\rm{GeV}$.
This possibility appears to us difficult to justify since 
it is more reasonable to expect the virtuality of the 
glueball, of order $m_{\eta'}^2 $, 
rather than its mass to be the relevant scale in such a process.

Our result for the Standard Model $b \to s g \eta'$
branching ratio casts doubt on the gluon fusion mechanism proposed in Ref.~\cite{du} 
as a possible explanation of the large exclusive $B \to \eta' K$ rate.
In this process the gluon from the $b \to s g \eta'$ transition is highly virtual
and is absorbed by the spectator. 
The authors of~\cite{du} calculate the exclusive rate using a perturbative QCD
method which we believe is unreliable in this case, given that a large contribution
in their approach arises for $q^2 \lsim 1 GeV^2 $.
In fact, using essentially the same $gg\eta'$ form factor 
as in Eq.~\ref{eq:formfactor}
they obtain an exclusive rate which is at least as large as
the corresponding inclusive $b \to s g \eta'$ rate, whereas one naturally
expects it to be significantly smaller.

Thus far we have considered short-distance mechanisms 
for $\eta'$ production in $B \to X_{sg}$ decays.  In particular, 
in both the factorization model 
and $b \to \eta' sg $ subprocesses, fast $\eta'$ production 
is associated with large gluon virtualities, i.e, $q^2 >> 1~\rm{GeV}^2 $.
It is important to note that there could also be important long-distance
quark and gluon fragmentation contributions.
As in the case of kaon production, we obtain an order of
magnitude estimate of soft $q \bar q$ popping contributions
using the $b \to sg$ Monte Carlo.  
Taking $\BR(b \to sg )\approx 10\%$, JETSET 7.4 with DELPHI tunings 
gives
$\BR(B \to \eta' X_s ) \sim 1\times 10^{-4}$ 
for $p_{\eta' } > 2$ GeV.  
As always in soft fragmentation processes, the large energy release leads 
predominantly to high multiplicity final states resulting
in a soft $\eta'$ spectrum, or a recoil spectrum
that would be consistent with observation.
In the Standard Model the $\eta' $ yield from quark fragmentation 
must be an order of magnitude smaller, i.e., $\BR(B \to \eta' X_s ) \sim 10^{-5}$,
since the underlying $b \to s \bar q q$ 
branching ratio is $\sim 1\%$.
It is worth mentioning that JETSET 7.4 with similar tunings
gives a reasonable description of both the $\eta'$ rate and momentum spectrum
observed in $Z$ decays~\cite{ALEPHQCD,OPALgamma}.  
Again, we stress that the $B \to X_{sg}$ Monte Carlo only provides very rough 
order of magnitude estimates.  For example, interference between contributions proceeding via the 
$u \bar u$, $d \bar d$, and $s \bar s$ components of the $\eta'$
can not be included in this approach.

Given the significant glue content of the $\eta'$
one might expect that 
soft ``$gg$ popping" (long-distance gluon fragmentation) is as 
important as soft $q \bar q$ popping in chromomagnetic $B \to X_{sg} $
transitions.
Because this is a long-distance process there would be no 
$q^2$ suppression, unlike in the case of the decays $b \to s g \eta'$.
Again, the large energy release
would lead predominantly to soft $\eta'$,
or a hard recoil spectrum.
Unfortunately, there is even less guidance from experiment in 
this case than in the case of quark fragmentation.
The similarity of the $\eta_c \to gg$ mediated 
decay rates~\cite{pdg96}\footnote{
The ratio
of the first two branching ratios~\cite{markiiietac} is $.76  \pm .3$.},
$\BR(\eta_c \to \eta' \pi \pi) = .041 \pm .017$,
$\BR(\eta_c \to \eta \pi \pi ) = .049 \pm .018$, and 
$\BR(\eta_c \to K \overline{K} \pi) = .066 \pm .018$,  
suggests that the $\eta' \pi \pi $ mode, like the other two,
proceeds via the intermediate transition 
$gg \to \bar q q \bar q' q'$.
In particular, there is no evidence for 
$\eta'$ formation out of gluons in this decay. 
On the other hand, we know that soft fragmentation
processes should lead mainly to high multiplicity final states
for which no data is currently available.   
In any case, long-distance gluonic $\eta' $ yields from
Standard Model $b \to sg^{(*)}$ decays should be at least an order of magnitude
smaller than those
obtained with $\BR(b \to sg) \approx 10\%$,
particularly since the Standard Model transition is dominated by the
chromoelectric form factor, which unlike the chromomagnetic form factor is not peaked
at low $q^2$.

In Ref.~\cite{fritzsch} an {\it ad hoc} effective Hamiltonian 
\beq H_{eff} = a \alpha_s G_F \bar{s}_L b_R G_{\mu \nu} \tilde{G}^{\mu \nu} \label{eq:HGG}
\eeq
was proposed to model the long-distance gluon fragmentation contributions 
to $\eta'$ production in the Standard Model.  
In this scenario the $\eta' $ is formed out of 
$G \tilde{G}$
and it is assumed that a large enough coefficient 
$a$ is generated to account for the observed $\eta'$
yield.  In addition 
to leading to a quasi two-body recoil spectrum
in conflict with experiment and with  
the expectation from soft fragmentation processes, 
this proposal suffers from another serious drawback. From the vacuum to $\eta'$ matrix
element of $G \tilde G$~\cite{tytgat},
the authors of Ref.~\cite{houhe} conclude that the central value of 
the $\eta'$ yield in Eq.~\ref{eq:cleoinc}
would correspond to $a \approx .015$ ${\rm GeV}^{-1}$.  It is then a simple matter to show
that Eq.~\ref{eq:HGG} would imply a fantastically large inclusive rate
$\BR(b \to s gg) \approx 10 - 50 \%$, where the range is obtained by varying
the scale at which $\alpha_s$  
is evaluated from $m_b$ to $m_{\eta'}$. The latter is the scale entering
the $G \tilde G$ matrix element, or the scale at which $a$ is determined.
In other words, a large $\eta'$
yield would be associated with a non-perturbative enhancement of 
the $b \to s\, {\rm glue} $ rate in the Standard Model 
of more than an order of magnitude, which is 
not plausible.  Conversely, it is unreasonable to expect a
contribution to the $B \to \eta' X_s$
branching ratio in this framework
much larger than $10^{-5}$.

Table~\ref{tab:three} summarizes the orders of magnitude of the various 
contributions to $\BR(B \to \eta' X_s )$ for $2.0 < p_{\eta'} < 2.7$ GeV, 
in the Standard Model and in models with
$\BR(b \to sg) \approx 10\%$. If long-distance gluon fragmentation is as important as 
quark fragmentation then there is an additional contribution of 
order $10^{-5}$ in the Standard Model and 
of order $10^{-4}$ for enhanced $b \to sg$.
The quasi two-body $b \to \eta' X_s$ 
recoil spectrum is peaked at low recoil mass, e.g., $m_{X_s}\sim 1.4$ GeV.
The shapes of the remaining recoil spectra are consistent with experiment.
It is clear from the Table that the existence of a cocktail solution for
the observed signal 
is very likely
in models with enhanced $b \to sg$.  However, the situation is more problematic in the 
Standard Model.  Apart from the $\approx 10^{-4}$ 
branching ratio contribution due to decays of intermediate charmonia,
all other contributions with consistent recoil spectra are of order 
$10^{-5}$.  Should the measured branching ratio remain near the current value,
our results suggest that a Standard Model explanation will require a
novel non-perturbative mechanism 
leading to order of magnitude enhancement of anomaly mediated 
processes. 

\begin{table}[t]
\caption{
Orders of magnitude for various contributions to $\BR(B \to \eta' X_s ) $
($2.0 < p_{\eta'} < 2.7$ GeV)
in the Standard Model and in models with $\BR(b \to sg) \approx 10\%$.
The entries, from top to bottom, correspond to decays of intermediate charmonia,
the quasi two-body and quasi three-body factorization 
model contributions, the anomaly induced $b \to s g \eta'$ decay, 
and long-distance quark fragmentation, as explained in the text.\label{tab:three}}
\vspace{0.4cm} 
\begin{center}
\begin{tabular} {| c | c | c |} \hline
Process & Standard Model & Enhanced $b \to sg$  
\\
\hline
$\overline{B} \to (c \bar c) X_s , (c \bar c) \to \eta' X$~\cite{atwoodsoni}  &  
$\approx 1.1 \times 10^{-4} $  & $\approx 1.1 \times 10^{-4}$   \\
$b \to \eta' s $ & $ 10^{-4} $ & $ 10^{-4} $  \\
$\overline{B} \to \eta' s \bar q $ &  $ 10^{-5}$      &    $ 5 \times 10^{-5} $ \\
$b \to s g \eta' $ & $ 10^{-5} $ & $ 10^{-4} $  \\
LD quark fragmentation & $ 10^{-5} $ & $ 10^{-4} $\\ 
\hline
\end{tabular}
\end{center}
\end{table}

\section{Conclusion}

Before concluding I would like to mention three novel features of
radiative $B$ decays which can arise in models with enhanced $b \to sg$: 
\begin{itemize}
\item Large direct CP asymmetries in the inclusive decays $B \to X_s \gamma$
vs. $\overline{B} \to X_{\bar s} \gamma$
of 10\% - 50\% are possible~\cite{kaganneubertCP}.
In the Standard Model the asymmetry is only of order 1\% due to a 
combination of CKM and GIM suppression, both of which can be lifted 
in New Physics scenarios with additional contributions to the dipole operators
containing new weak phases.  
The asymmetries provide
a unique probe of models with enhanced $b \to sg$ since they 
are particularly sensitive to
interference of the next-to-leading order one-loop diagram containing a chromomagnetic
dipole operator insertion (which generates a strong phase)
and the tree-level diagram for the electromagnetic dipole operator. 
\item New chromomagnetic dipole operator mediated graphs in which 
the spectator quark radiates a photon
can lead to large isospin violation~\cite{petrov2} in $B \to K^* \gamma $.
In particular, rate asymmetries between the $K^{*-} \gamma$ and $K^{*0} \gamma$ 
final states could exceed 50\%, compared to only a few \% in the standard model.
\item In models in which $b \to d g$ is also strongly enhanced, e.g., in
association with generation of $V_{ub}$, $b \to d \gamma $ 
is necessarily enhanced 
so that $\BR(B \to \rho \gamma,\omega \gamma)$ could be an order of magnitude
larger than in the Standard Model~\cite{kagan}. 
\end{itemize}

To summarize, there are two potential hints for enhanced 
$b \to sg$ which are essentially experimental, 
a $2 \sigma$ deficit in charm counting and a $3 \sigma $ deficit in
kaon counting.
Significantly improved precision in charm and kaon counting
will require a reduction in uncertainties in the
absolute $D /D_s $ branching scales.  Improved precision in kaon counting
will also require an experimental determination of the amount of $s \bar s$
popping in $B$ decays
via future measurements of 
$\BR(\overline{B} \to D K \overline{K} X)$ and 
$\BR(\overline{B} \to D_s \overline{K} X) $. 
There are also well known hints from comparison of next-to-leading order Standard Model predictions
for the charm multiplicity and semileptonic branching ratio with measurements
at the $\Upsilon (4S)$.  Unfortunately, improved theoretical precision 
poses a difficult challenge in the near future.

A promising direct search strategy is the search for 
high momentum kaon excesses in $B$ decays, e.g., $p_K \gsim 1.8~GeV$. 
A JETSET analysis indicates that for $\BR(b \to sg) \sim 10\%$ the 
corresponding $B \to KX$ branching ratios are of order $10^{-3}$.
The Standard Model background at these momenta is of same order
and is dominated by kaon yields from intermediate $D^0 /D^+ $ decays,
which can be determined experimentally.
Remarkably, enhanced $b \to sg$ can evade all existing rare $B$ decay
constraints.  We have also seen that new weak phases in 
the chromomagnetic dipole operator coefficients can lead to large $CP$ violation
in hadronic rare decays.  
For example, direct $CP$ asymmetries in 
$B^\pm \to \phi K^\pm $ and $B^\pm \to K^0 \pi^\pm $
could be as large as $10\% - 50\%$. However, additional 
flavor $SU(3)$ tests will be required in order to 
distinguish such contributions from 
potentially large Standard Model asymmetries induced by soft final
state rescattering.  Furthermore, such large New Physics 
contributions are not restricted to 
models with enhanced $b \to sg$.  
In contrast, similarly large direct CP asymmetries in
radiative $B \to X_s \gamma$ decays would provide unambiguous evidence for
strongly enhanced chromomagnetic dipole operators.

The large $B \to \eta' K$ branching ratios measured by CLEO
can be accounted for in the Standard Model in the factorization approach.
Low values of $m_s$ and particular ranges of parameters
for non-factorizable contributions are favored.
Nevertheless, enhanced $b \to sg$ can lead to significant enhancement of 
the factorization model rates.  A Standard Model explanation for the
large inclusive $B \to \eta' X_s $ rate measured by CLEO is 
more challenging.  It seems that a novel non-perturbative mechanism 
is required which would enhance the gluon anomaly mediated contributions by an order of
magnitude.  In contrast, enhanced $b \to sg$ can increase
the gluon anomaly, long-distance quark fragmentation,
and quasi three-body factorization model contributions by an order of magnitude each
so that the existence of a cocktail solution becomes very likely.  The 
inclusive $\eta'$ rate may therefore be providing us with another hint 
for enhanced $b \to sg$, or TeV scale flavor dynamics.

\section*{Acknowledgments}
Much of the work presented here was done in collaborations with Antonio Perez,
Alexey Petrov, and Johan Rathsman, for which I am grateful.  It is
also a pleasure to thank Adam Falk, Matthias Neubert and Yossi Nir for 
very pleasant collaborations on relevant material.  I would also  
like to thank Marco Battaglia, Mark Convery, Su Dong, Sue Gardner, Keh-Fei Liu,
Jim Smith, Mike Sokoloff, and Arkady Vainshtein for useful discussions.
This work was supported by the United States Department of Energy under Grant
No. DE-FG02-84ER40153.

\end{document}